\documentclass{moriond}

\newcommand\Tstrut{\rule{0pt}{2.6ex}}         
\newcommand\Bstrut{\rule[-0.9ex]{0pt}{0pt}}   
\usepackage{amsmath}
\usepackage{amssymb}
\def\Journal#1#2#3#4{{#1} {\bf #2}, #3 (#4)}

\def\PRD{{\em Phys. Rev.} D}

\def\be{\begin{equation}}
\def\ee{\end{equation}}
\def\bea{\begin{eqnarray}}
\def\eea{\end{eqnarray}}

\newcommand{\Photo}{\includegraphics[height=35mm]{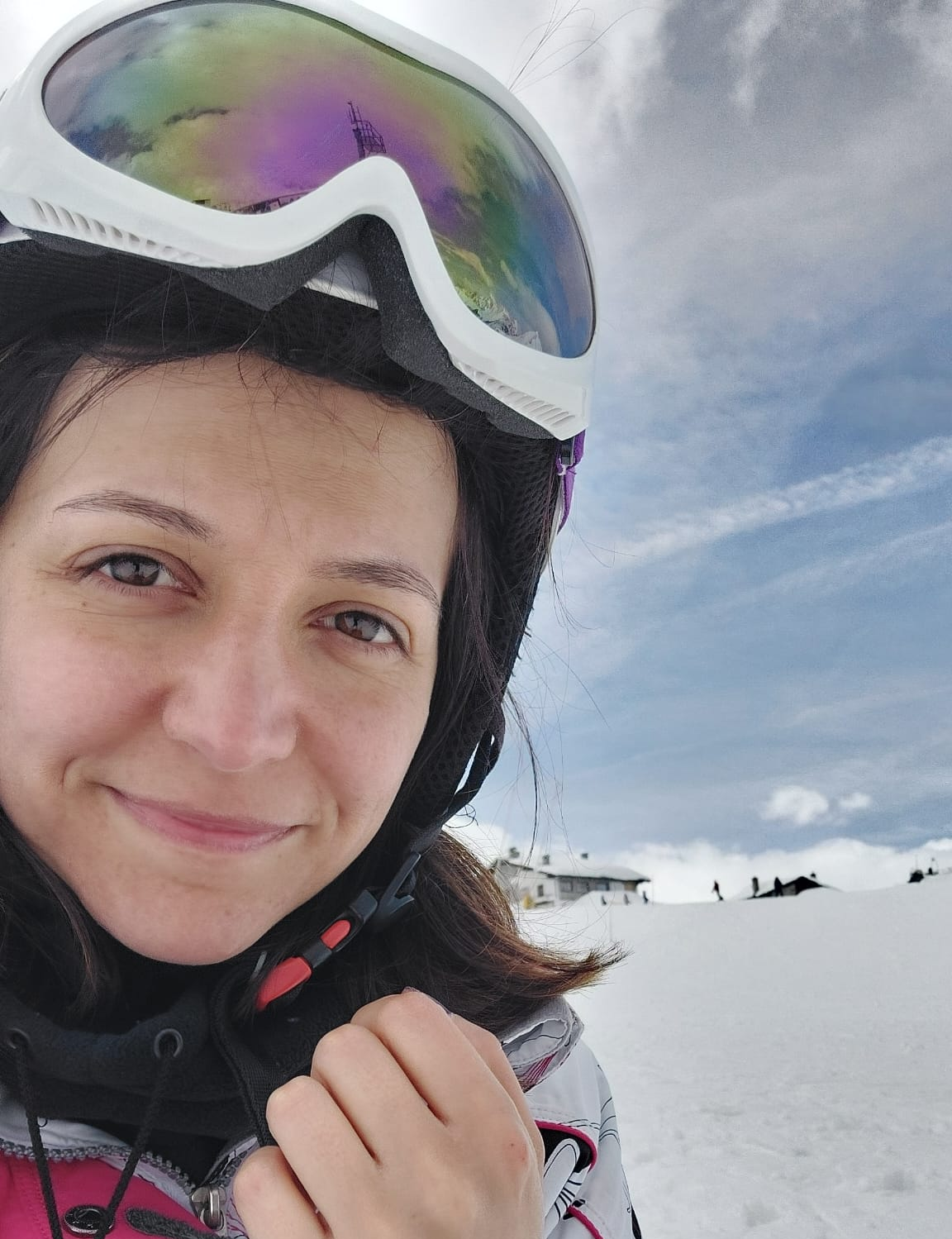}}

\begin{document}
\vspace*{4cm}
\title{A model-independent treatment of cosmic ladder calibration and $\Omega_k$ measurement through low-$z$ observations}


\author{A. Favale$^{1,2}$~\footnote{Speaker, arianna.favale@roma2.infn.it}
          ,
          A. G\'omez-Valent$^{2}$
          and
          M. Migliaccio$^{1}$
          }

   \address{$^{1}$ Università di Roma Tor Vergata \& INFN Sez. di Roma 2, Via della Ricerca Scientifica,1, 00133 Roma, Italy\\  
            $^{2}$ Departament de Física Quàntica i Astrofísica, and Institute of Cosmos Sciences, Universitat de Barcelona,
Av. Diagonal 647, 08028 Barcelona, Spain
            }

\maketitle\abstracts{Looking at the well-known Hubble tension as a tension in the calibrators of the cosmic distance ladder, i.e. the absolute magnitude $M$ of standard candles such as supernovae of Type Ia (SNIa) and the standard ruler represented by the comoving sound horizon at the baryon-drag epoch, $r_d$, we propose a model-independent method to measure these distance calibrators independently from the cosmic microwave background and the first rungs of the direct distance ladder. To do so, we leverage state-of-the-art data on cosmic chronometers (CCH), SNIa and baryon acoustic oscillations (BAO) from various galaxy surveys. Taking advantage of the Gaussian Processes Bayesian technique, we reconstruct $M(z)$, $\Omega_k(z)$ and $r_d(z)$ at $z\lesssim2$ and check that no significant statistical evolution is preferred at 68\% C.L. This allows us to treat them as constants and constrain them assuming the metric description of gravity, the cosmological principle and the validity of CCH as reliable cosmic clocks, and SNIa and BAO as optimal standard candles and standard rulers, respectively, but otherwise in a model-independent way. We obtain: $\Omega_k=-0.07^{+0.12}_{-0.15}$, $M=(-19.314^{+0.086}_{-0.108})$ mag and $r_d=(142.3\pm 5.3)$ Mpc. At present, the uncertainties derived are still too large to arbitrate the tension but this is bound to change in the near future with the advent of upcoming surveys and data.
}

\section{The $H_0$ tension as a tension in the calibrators of the cosmic distance ladders}
The discrepancy between the model-independent measurement of $H_0$ obtained by SH0ES~\cite{riess} calibrating supernovae of Type Ia (SNIa) in the local distance ladder and the $\Lambda$CDM-based result inferred by \textit{Planck}'s observations of the cosmic microwave background (CMB)~\cite{planck} stands as the most statistically significant tension in modern cosmology. When CMB measurements are used to get a model-dependent estimate of the sound horizon at the baryon-drag epoch, $r_d$, cosmic distances can be extracted from data on baryon acoustic oscillations (BAO) and the inverse distance ladder can be applied providing an alternative approach to measure $H_0$.
Depending on the data set used in the analysis, values of $H_0$ estimated independently by both late and early time probes through the direct and inverse distance ladders are in tension at $\approx 4\sigma-6 \sigma$ level (see, e.g., \cite{edv21} for a dedicated review).
In light of this, it becomes evident that the $H_0$ tension can be reframed as a tension in the calibrators of the direct and inverse distance ladders, namely the absolute magnitude of SNIa, $M$, and $r_d$.
By determining these two quantities independently from the CMB and the first rungs of the direct distance ladder, we can have a means to cross-check the results obtained with the aforementioned standard methods and also to assess the viability of models beyond $\Lambda$CDM that have been widely proposed to alleviate the tension.

\section{Agnostic \& data-driven analysis to calibrate the ladders and measure the curvature of the universe}

\begin{figure}[t!]
\centering
\includegraphics[scale=0.17]{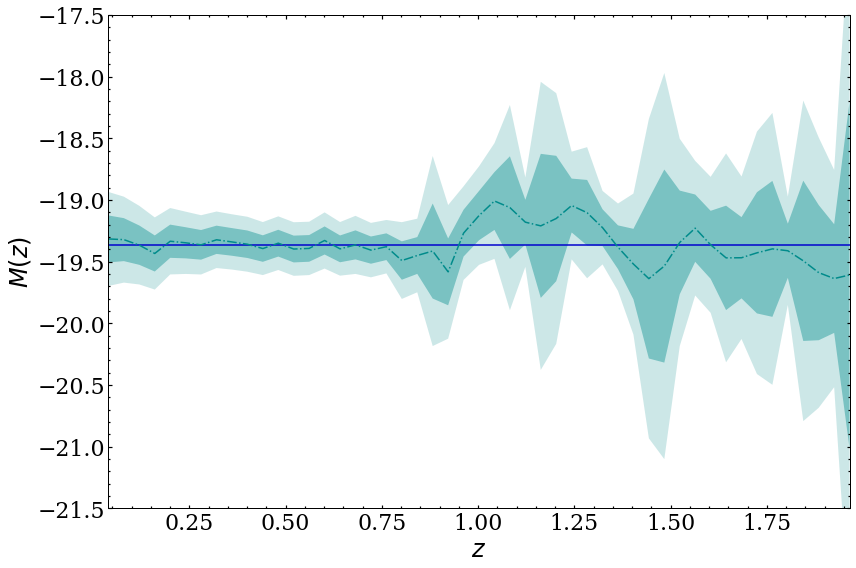}
\includegraphics[scale=0.17]{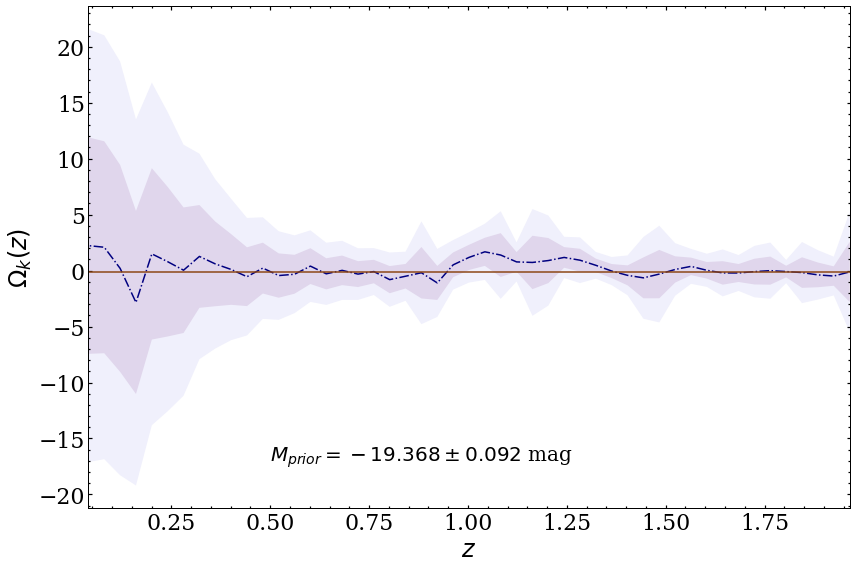}
\includegraphics[scale=0.17]{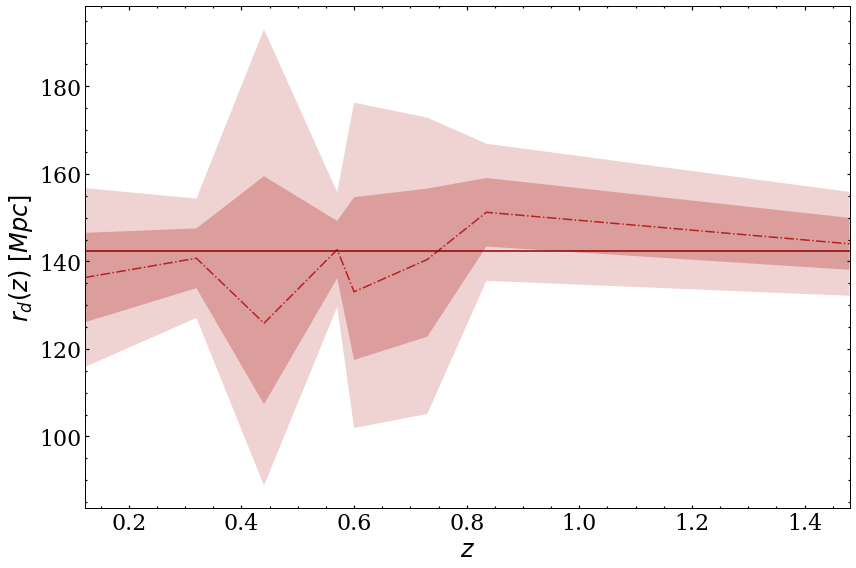}
\caption{Reconstructed shapes of $M(z)$, $\Omega_k(z)$ and $r_d(z)$ at 68\% and 95\% C.L., showing that all these quantities are compatible with a constant value (i.e. the weighted average in solid line) at $\sim$68\% C.L.}
\label{fig:3plots}
\end{figure}

State-of-art data at $z\lesssim2$ can be employed together with advanced statistical techniques to reconstruct the expansion history of the universe in a quite model-independent way. This allows to test some of the basic assumptions that lie behind the $\Lambda$CDM model, which are usually taken for granted in other works, i.e. the constancy of the absolute magnitude of SNIa throughout the cosmic history and the homogeneity of the universe at large scales. Departures from these assumptions would be a hint of new physics beyond the standard model, which could be a manifestation of the breaking of the cosmological principle or of modified gravity effects, but it could also point to unaccounted systematics in the data. Therefore, in these proceedings, which are based on the work~\cite{favale23}, we reconstruct both $M$ and the curvature parameter of the universe, $\Omega_k$, as a function of the redshift to test their constancy in time. To do so, we employ \textit{Gaussian Processes} (GP) \cite{gp}, which is a Bayesian tool designed to perform data-driven reconstructions from Gaussian distributed data. Not needing to specify an underlying model, the GP provides agnostic estimates of functions of interest. A kernel function, which in most cases depends on only two hyperparameters, has to be specified so that one can track the covariance between points at which there are no data. We validate our findings against different kernel functions and by taking into account the propagation of the uncertainties of their hyperparameters. In particular, we apply GP employing observations free from the main drivers of the current tensions: (\textit{i}) measurements of $H(z)$ from cosmic chronometers (CCH) obtained using massive and passively-evolving galaxies and the diﬀerential-age technique. We employ them as calibrators of the cosmic ladders; (\textit{ii}) apparent magnitudes of SNIa, $m(z)$, from the Pantheon+ compilation~\cite{pant}; (\textit{iii}) anisotropic angular ($D_V(z)/r_d$, $D_M(z)/r_d$) and radial ($r_{d}H(z)$) BAO data from different galaxy surveys (6dFGS, BOSS, eBOSS, WiggleZ, DES Y3). We duly account for the available statistical and systematics errors for all the data sets (see Tables 1 and 2 of~\cite{favale23}).

\noindent In a FLRW universe, the luminosity distance can be written as
\begin{equation}\label{eq:dL}
d_L(z) = \frac{c(1+z)}{\sqrt{\Omega_k H_0^2}}\sinh\left(\sqrt{\Omega_k H_0^2}\int_0^z\frac{dz^\prime}{H(z^\prime)}\right)\,,
\end{equation}%
with the sign of $\Omega_k$ set by the spatial geometry of the universe. Using this expression one is able to evaluate the SNIa apparent magnitude as
\begin{equation}\label{eq:mfinal}
    m(z) = M +25+5\log_{10}\left(\frac{d_L(z)}{1\,{\rm Mpc}}\right)\,.
\end{equation}%

\begin{figure}[t!]
\centering
\includegraphics[scale=0.18]{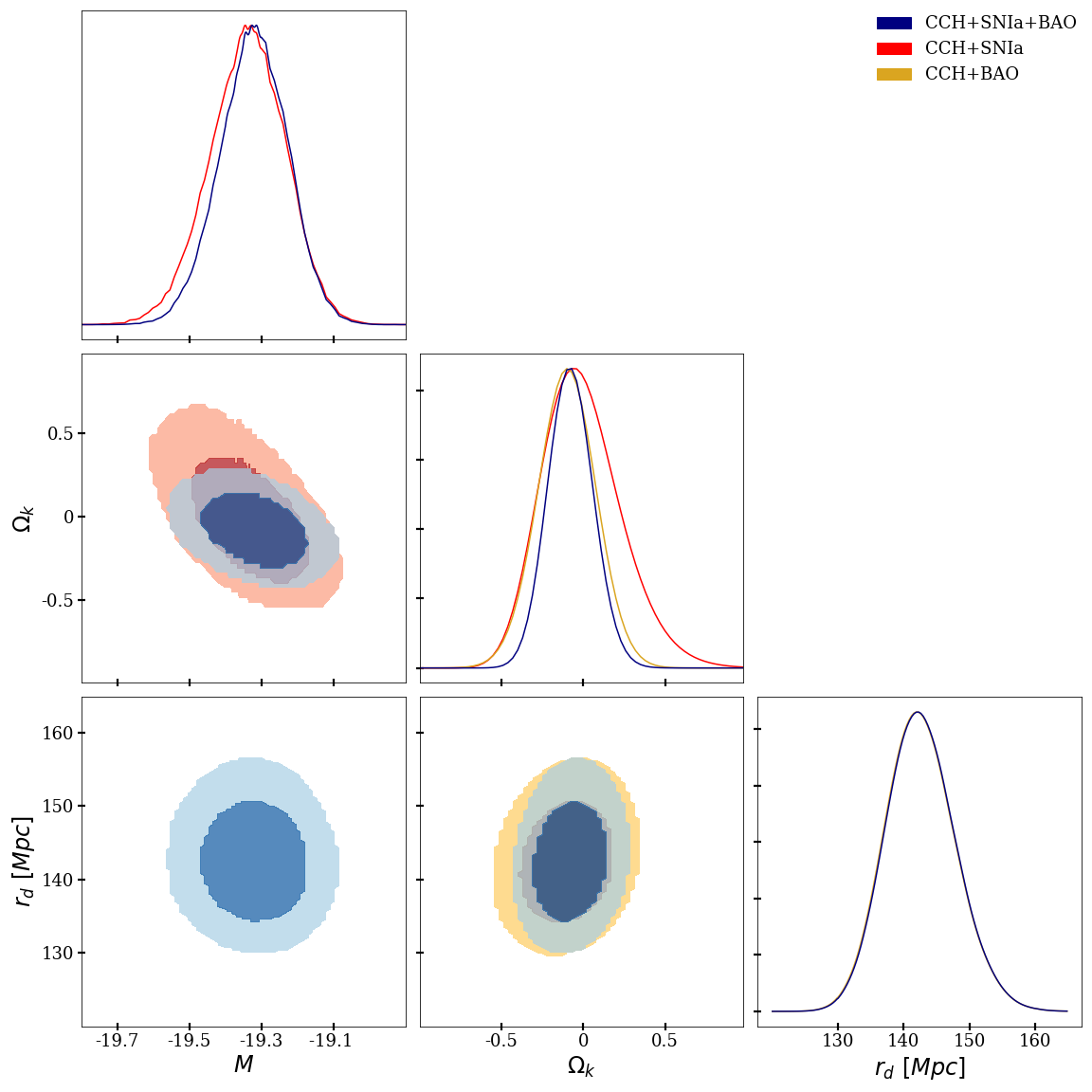}
\includegraphics[scale=0.18]{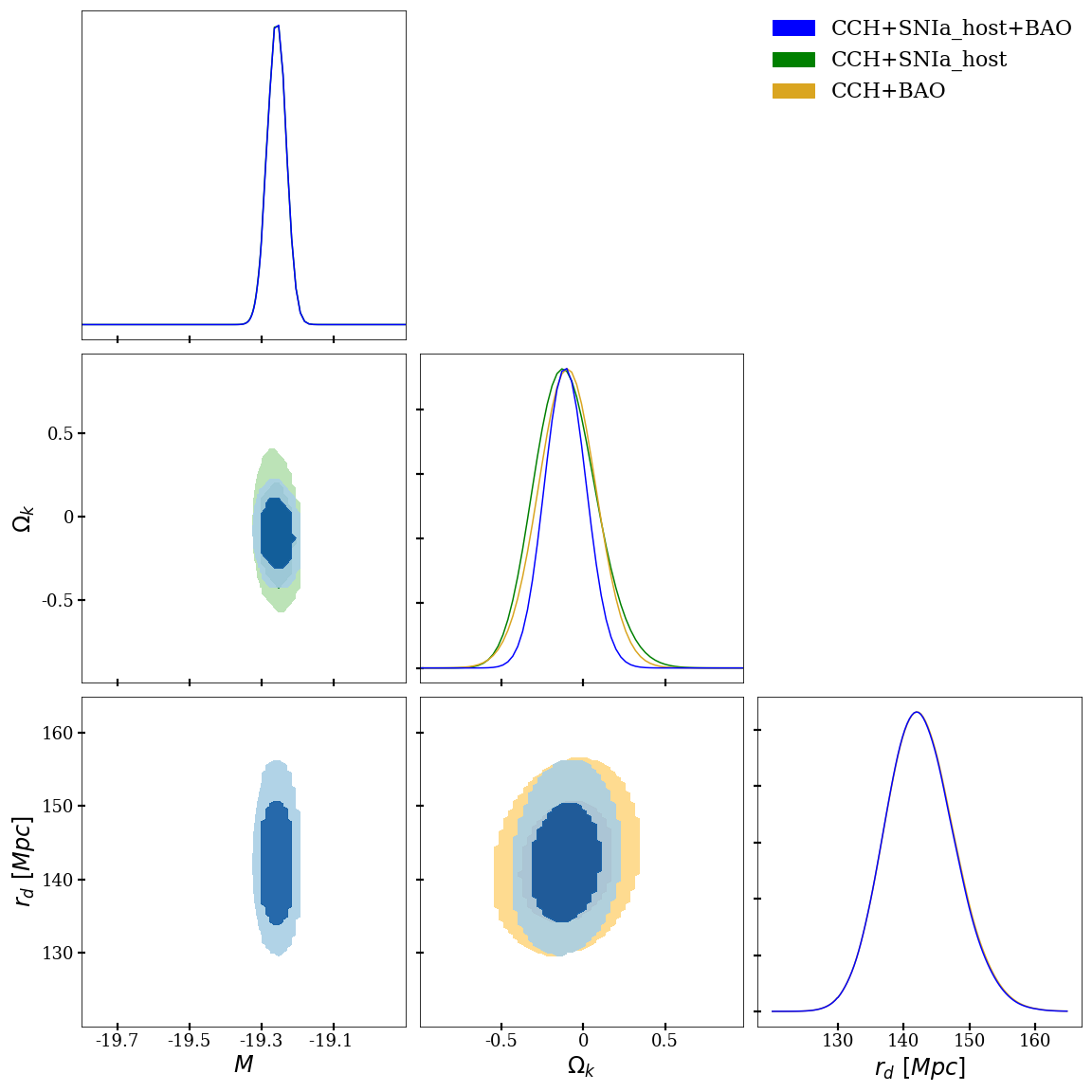}
\caption{1D posterior probability distributions and 2D contours at 68\% and 95\% C.L. for the joint analyses in all the planes of the parameter space ($M, \Omega_k, r_d$). The two plots differ for the SNIa data employed: on the left-hand side, we exclude from the analysis the SNIa calibrated in the Cepheid host galaxies which, as expected, have an important impact on the posterior of the absolute magnitude $M$ (cf. right-hand side plot and Table \ref{tab:joint}).}\label{fig:joint}
\end{figure}%
\noindent In practice, with GP we reconstruct the shape of $H(z)$ from CCH and the one of $m(z)$ from SNIa and use these results to obtain $\Omega_k(z)$ and $M(z)$. The $H(z)$ reconstruction can also be used to test the internal consistency of the BAO data set in order to ensure that there is no significant statistical tension between the values of $r_d$ obtained from each BAO measurement.

\begin{table}[t!]
    \centering
    \caption{Constraints on $M$ (in mag), $\Omega_k$ and $r_d$ (in Mpc) at 68\% C.L. obtained from the joint analyses with and without the inclusion of the SNIa calibrated in the Cepheid host galaxies (SNIa\_host and SNIa, respectively). The CCH+BAO results are reported only once since they are not affected by the choice of the SNIa sample.}
    \label{tab:joint}
    \vspace{0.4cm}
    \begin{center}
    \begin{tabular}{|c|c|c|c|c|c|}
    \hline 
         & \footnotesize{CCH+SNIa} & \footnotesize{CCH+BAO} & \footnotesize{CCH+SNIa+BAO} & \footnotesize{CCH+SNIa\_host} & \footnotesize{CCH+SNIa\_host+BAO}\Bstrut\\ \hline \Tstrut  
        M & $-19.344^{+0.116}_{-0.090}$ &  & $-19.314^{+0.086}_{-0.108}$ & $-19.252^{+0.024}_{-0.036}$  & $-19.252^{+0.024}_{-0.036}$  \Bstrut\\ \hline \Tstrut
         $\Omega_k$ & $-0.07^{+0.27}_{-0.21}$ & $-0.10\pm0.18$ & $-0.07^{+0.12}_{-0.15}$ & $-0.13^{+0.18}_{-0.21}$ & $-0.10^{+0.12}_{-0.15}$\Bstrut\\ \hline \Tstrut 
         $r_d$ & & $141.9^{+5.6}_{-4.9}$& $142.3\pm5.3$ & & $141.9^{+5.6}_{-4.9}$\Bstrut\\ \hline  
         
    \end{tabular}
    \end{center}
\end{table}%
\noindent We find that $M$, $\Omega_k$ and $r_d$ are all compatible with a constant value at $\sim68\%$ C.L. (see Fig. \ref{fig:3plots}), thus no evolution is preferred by current data. We are therefore allowed to perform a joint analysis in the full parameter space of $(M, \Omega_k, r_d)$ by treating them simply as constants. In this way, we can obtain consistent constraints for the curvature parameter and the calibrators of the distance ladder. We use CCH to calibrate the standard candles and the standard rulers and obtain constraints in the planes $(M,\Omega_k)$ and $(r_d,\Omega_k)$ using CCH+SNIa and CCH+BAO, respectively. Since SNIa and BAO data are independent, we can finally perform a CCH+SNIa+BAO joint analysis by means of the evaluation of $\chi^{2}(M,\Omega_k, r_d) = \chi^{2}(M,\Omega_k)+\chi^{2}(\Omega_k, r_d)$. This last analysis leads in particular to the tightest constraints achieved within this agnostic approach. They read: $\Omega_k=-0.07^{+0.12}_{-0.15}$, $M=(-19.314^{+0.086}_{-0.108})$ mag and $r_d=(142.3\pm 5.3)$ Mpc. The calibration of the standard ruler leads to a value which is $\sim1\sigma$ smaller than the $\Lambda$CDM value preferred by Planck. On the other hand, the constraints on parameters like $\Omega_k$ offer valuable insights into the early universe and the inflationary period. We find a mild preference for a closed universe, although compatible with the flatness assumption at 1$\sigma$ C.L.
Additionally, we test the impact of the inclusion of the SNIa located in the Cepheid host galaxies, i.e. those calibrated by SH0ES in the second rung of the ladder used to estimate $H_0$. We find that, whereas affecting very mildly the constraints for $r_d$ and $\Omega_k$, it completely dominates the result for $M$, making its central value and uncertainties biased towards those measured by SH0ES, $M^{R22} = -19.253\pm0.027$ mag \cite{riess}. See triangle plots in Fig. \ref{fig:joint} and the derived constraints in Table \ref{tab:joint}.


\section{Conclusions and outlook}

\begin{figure}[t!]
\centering
\includegraphics[scale=0.17]{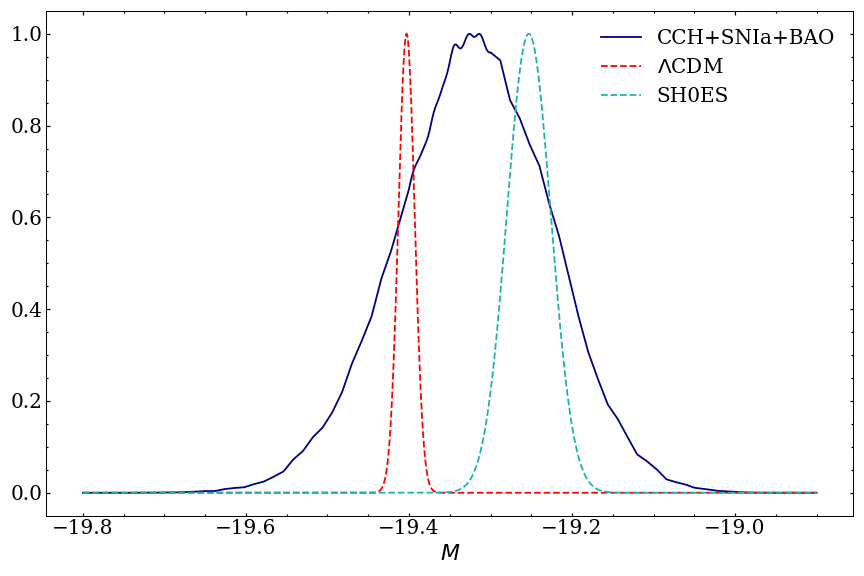}
\includegraphics[scale=0.17]{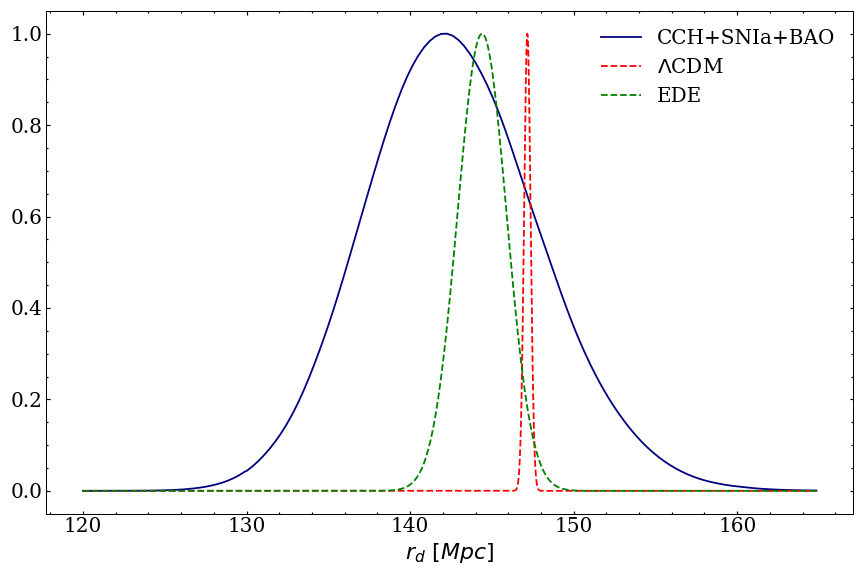}
\includegraphics[scale=0.17]{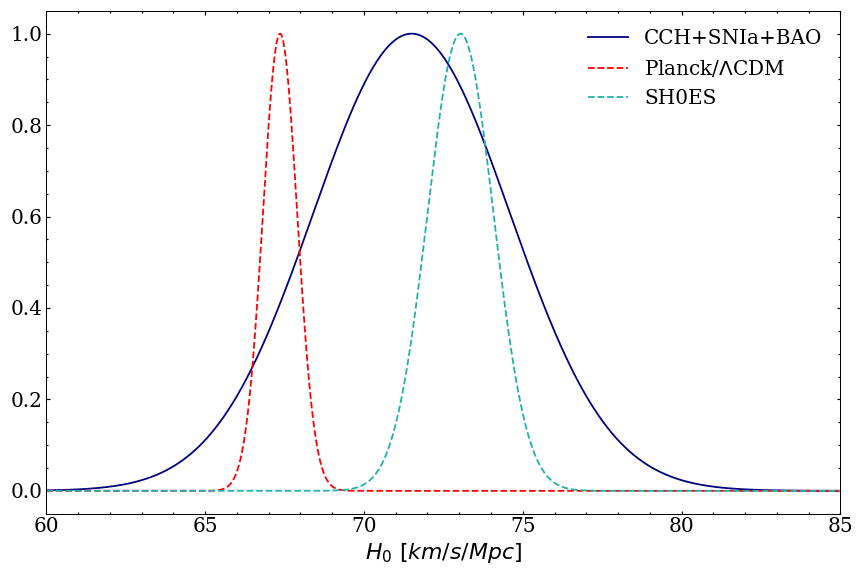}
\caption{Model-agnostic constraints on the calibrators of the distance ladder, $M$ and $r_d$, and on $H_0$, obtained employing the results of the CCH+SNIa+BAO analysis. We compare them with the constraints obtained in the context of the $\Lambda$CDM model ('$\Lambda$CDM') or early dark energy model ('EDE') from the fitting analysis with Planck2018+SNIa+BAO data \protect\cite{gv22}; the SH0ES result on $M$ and $H_0$ \protect\cite{riess}; the {\it Planck}/$\Lambda$CDM \protect\cite{planck} $H_0$ value.}
    \label{fig:results_comp}
\label{fig:joint}
\end{figure}%

Obtaining model-independent constraints on $\Omega_k$ and the calibrators of the distance ladders, $M$ and $r_d$, is of fundamental importance for a better characterization of the Hubble tension and the early universe. In our work, we have been able to do so by calibrating the ladders with cosmic chronometers. The uncertainties found with our method are still too large to arbitrate the $H_0$ tension (see Fig. \ref{fig:results_comp}). Indeed, employing the CCH+SNIa+BAO result on $M$ and the SNIa in the Hubble flow ($0.02<z<0.15$), and using the cosmographic expansion of $d_L$\footnote{From the SNIa in the Hubble flow we constrain the deceleration parameter as $q_0 = -0.44\pm0.19$.},
\begin{equation}
d_L(z)=\frac{cz}{H_0}\left[1+\frac{z}{2}\left(1-q_0\right)\right]+\mathcal{O}(z^3)\,,
\end{equation}%
we find $H_0=(71.5\pm 3.1)$ km/s/Mpc. However, the constraining power of our method is bound to improve thanks to forthcoming surveys (e.g., Vera C. Rubin LSST, Euclid, DESI) expected to provide high-quality data and increase the statistics especially at low $z$.
Moreover, this work can be naturally extended to calibrate other standardizable objects, like gamma-ray bursts, which can help us to build and extend the distance ladder beyond the region covered by SNIa and BAO ($z>2$) in a model-independent way (see, e.g., \cite{favale24}). This will contribute to a better  understanding of the universe dynamics in a region of redshifts which is currently poorly explored.

\section*{Acknowledgments}
A. Favale acknowledges the support of the Department of Physics of the University of Rome La Sapienza, University of Rome Tor Vergata and the INFN project “In-Dark”, which has made possible her participation in the 58th Rencontres de Moriond.

\end{document}